\documentclass[preprint]{aastex}

\shortauthors{Kawaler \& Hostler}
\shorttitle{sdB Rotation and Asteroseismology}

\begin{document}

\title{Internal rotation of subdwarf B stars: limiting cases and
asteroseismological consequences}

\author{Steven D. Kawaler \altaffilmark{1,2}}
\author{Shelbi R. Hostler \altaffilmark{1,3}}

\altaffiltext{1}{Department of Physics and Astronomy, Iowa State University,
Ames, IA  50011 USA} 
\altaffiltext{2}{e-mail: sdk@iastate.edu}
\altaffiltext{3}{current address: Submillimeter Array, 645 North A'ohoku Place
Hilo, HI 96720}

\begin{abstract}

Observations of the rotation rates of horizontal branch (HB) stars
show puzzling systematics.  In particular, cooler HB stars often show 
rapid rotation (with velocities in excess of 10 km/s), while hotter HB stars
(those with $T_{\rm eff}$ in excess of 11,000K) typically show much
smaller rotation velocities of 8 km/s or less.  Simple models of angular
momentum evolution of stars from the main sequence through the red giant
branch fail to explain these effects.  Assumption of solid body rotation
throughout rotate much too slowly.  On the other hand, assuming local 
conservation of angular momentum, but with solid body rotation
in the convective regions, produces HB models also rotate too slowly at all 
$T_{\rm eff}$ -- but
preserve a rapidly rotating core.   In these cases, the observed angular 
velocities of HB stars require that some of the angular momentum stored in the 
core reaches the surface.  Models that assume constant specific 
angular momentum in the surface convection zone have faster rotation 
in the envelope; while the predictions of these models match the observed 
rotation rates of the cooler HB stars, hot HB stars rotate more slowly than
the models.  Thus, while there is not as yet a coherent explanation of the trends 
of rotation on the HB, evolutionary models in all cases preserve 
a rapidly rotating core.  To test the idea that HB stars contain such a core, 
one can appeal to detailed computations of trace element abundences and 
rotational mixing.  However, a more direct probe is available to test these 
limiting cases of angular momentum evolution.  Some of the hottest horizontal 
branch stars are members of the pulsating sdB class.  They frequently show rich 
pulsation spectra characteristic of nonradially pulsating stars.  Thus their 
pulsations probe  the {\it internal} rotation of these stars, and should show 
the effects of rapid rotation in their cores.  Using models of sdB stars that 
include angular momentum evolution, we explore this possibility and show that 
some of the sdB pulsators may indeed have rapidly rotating cores.

\end{abstract}


\section{Introduction}

\subsection{Rotation of horizontal branch stars}

First studied in detail by R. Peterson in the mid 1980s
\citep{pete83, pete85a, pete85b}, the relativley rapid rotation of some
horizontal branch stars remains an enigmatic feature of this phase of
evolution.  Rapid rotation was not seen in all horizontal branch (HB) stars, but
the work by \citet{Petetal95} on M13 suggested that rapid rotation was more 
common in cooler HB stars.  
Later observations by \citet{BehrM15} of HB stars in M15 and in
M13 \citep{BehrM13} further illuminate that there is a gap in the horizontal 
branches of M15 and M13, with two distinct rotation rates on either side.
At effective temperatures less than about 11000~K, HB stars frequently show
rapid rotation ($v \sin i$ $\approx $ 40 km/s) while no rotation rates above 
10 km/s appear in the hotter stars ($T_{\rm eff} > $ 11000 K). The work by
\citet{BehrM15, BehrM13} agreed with 
a previous survey of M13 HB stars completed by \citet{Petetal95}.  Further 
work by \citet{BehrGC} amplifies these results.  
In additon, a study 
of several other galactic globular clusters by \citet{Recetal02} and
\citet{Recetal04} found 
similar results for the rotation of cool and hot HB stars.  RR Lyra stars
appear to show rotation with velocities below 10 km/s
(eg \citep{petetal96}).  While 
it is possible that the pulsations  can drive a modest wind which in turn carries
away angular momentum from these pulsating stars, this pause in the trend of
rotation velocity with decreasing temperature is a mystery.

Age and metallicity are the primary factors in determining a cluster's 
features and placement in the H--R diagram, and (in principle) the structure of 
the cluster's HB.  Thus, if two clusters have similar metallicities and ages, one 
would expect them to show similar horizontal branch morphologies. But that is not 
the case; this is the infamous ``second parameter problem'' of globular cluster 
evolution. The rotation anomaly may couple to the other questions raised by HB 
morphology through the second parameter problem \citep{BehrM15,BehrM13,BehrGC} 
so an understanding of the rotational break could be important in understanding 
the fundamentals of stellar evolution.

Much of the work on horizontal branch stars naturally concentrates on stars
in globular clusters.  However, field stars that correspond to the post
helium core flash of low--mass stars can perhaps play an important role in
understanding the connection between rotation and HB morphology. 
For Population I stars, identification of post helium--flash objects is 
complicated by the paucity of stars in galactic clusters and distance 
determination for field stars.  However, one class of such stars, the subdwarf 
B stars, are very hot ($T_{\rm eff}\gtrsim 25,000K$ or more), and show high gravity 
($\log g \approx 5.3 - 6.1$).  Thus they are readily identified as blue point sources 
in multicolor surveys, and spectroscopy reveals their stellar nature.  They are 
the field analog to the bluest HB stars seen in clusters, and have very thin 
surface hydrogen layers.  

But these stars present their own enigmas - in particular the origin of
sdB stars.  Their origin as stars which have lost most but not all of their 
hydrogen envelope following RGB evolution is mysterious, as simple models of RGB 
evolution cannot easily explain their miniscule surface hydrogen layers.  Models 
proposed to explain their origin include interactions with a companion 
(eg. \citet{sandtaam, mengetal76, hanetal03})and even mass stripping by planetary 
companions \citep{SokHar00}.  The high
incidence of composite-spectrum binaries among the sdB stars is consistent with
these kinds of models.
Other scenarios involving single--star evolution include variations of mass loss 
efficiency on the RGB \citep{dcruz}, and subsequently experience the core helium flash
while on the helium--degenerate white dwarf cooling track (eg. \citet{brownetal01, 
swei97}).  
In some cases, the proposed origin scenario makes a clear prediction about
the rotational properties of sdB stars.  For example, those that involve
binary mergers or common envelope evolution should spin up the sdB product.

While there is no direct spectroscopic evidence because of the signature of 
chemical diffusion, the field sdB stars appear to be Population I objects.  Their 
kinematics suggest that, in general, they derive from a younger population, 
and therefore we assume in our modeling that their progenitors have masses 
of 1 $M_{\odot}$ or greater.

As far as their rotation, comparison with the results for Pop. II stars suggests 
that they all should be slow rotators, with $v \sin i \lesssim$ 10km/s.  
% Currently, there is limited data on their rotation rates (in part because 
% high S/N, high resolution  spectra are required).
Recently \citet{BehrField} measured rotation velocities of a sample of field HB 
stars showed that a large majority are slow rotators, but that a sample of field
HB stars do rotate faster.  The work by \citet{hebetal99} has shown that at least
one star, PG~1605, is a more rapid rotator -- as expected based on an asteroseismic 
analysis of the pulsations by \citet{Kaw98}.

\subsection{Models of rotation of HB stars}

Attempts to understand the rotation observations, and perhaps link the rotation
break to gaps in the horizontal branch or general HB morphology, illustrate 
the creativity of theorists.   The lack of a coherent answer to date is a testament 
to the difficulty of the problem.  Following up on a previous study of 
differential rotation in RGB stars \citep{Pinetal91}, a detailed investigation 
by \citet{SP2000} (hereafter SP2K) explored possible solutions based on single
star evolution with different assumptions about internal angular momentum 
transport.

As we follow the same general procedure here in our study of sdB stars, we 
describe the computations of SP2K in some detail. SP2K modeled rotation on the 
red giant branch (RGB) and its effect on HB rotation.  Starting with models that 
are rotating as solid bodies at the departure from the main sequence, SP2K followed 
evolution of the models through the helium flash and onto the HB.  They studied 
several different limiting cases of angular momentum transport, two of which we 
follow up on in our study of 
Population I hot HB stars.  Both cases conserve specific angular 
momentum (angular momentum per unit mass, $j$) in convectively stable 
regions.  In convective regions, convective mixing is assumed to 
produce either constant angular velocity in the convection zone
(Case B) or constant $j$ (Case D); in both cases, the angular momentum
contained within the convection zone is preserved.  These designations 
correspond to those used by SP2K.

For low mass Population II stars, SP2K concluded that observations of the rapidly 
rotating cool HB stars are best understood with models with rapidly rotating
cores that evolve from progenitors that have constant specific angular
momentum in convection zones (i.e. Case D).
Case D evolution produces models that can rotate at the rates seen in the cool HB
stars.  The slow rotation in the hotter HB stars requires some choking of
angular momentum transport by chemical composition gradients produced by
diffusion - such effects are seen in HB stars (and models) at temperatures
above 11,000K (SP2K).

An alternative to SP2K's analysis is given by \citet{SokHar00}, who investigated
interaction with a planetary companion as one suggestion for a "second
parameter."  \citet{SokHar00} argued that the progenitors of the
rapidly rotating HB stars have been spun up by planets or brown dwarfs,
masses approximately 5 times Jupiter's mass and at separations of about 2
AU.  While they state that these planetary interactions are the main
second parameter, they also admit that planets do not completely solve the
problem.   The models of \citet{SokHar00} rely on
horizontal branch angular momentum loss, not ineffective angular momentum
transport as in SP2K to explain the slower rotating blue horizontal branch stars. 

\subsection{Pulsating sdB stars - direct probes of internal rotation}

A subset of sdB stars are known to be multiperiodic pulsating stars.  As
nonradial modes must be present to explain the density of pulsation modes in
a single star, we may be able to probe the interiors of these stars through
asteroseismology.  Such work on the EC~14026 stars (or sdBV stars), has already
begun.  The general properties of the sdBV stars are reviewed from an observational 
perspective by \citet{Kilk02}; theoretical models of these stars are reviewed 
by \citet{Chaetal01}. 

Models of rotation in single HB stars indicate that these
stars should have rapidly rotating cores.  Similarly, sdB stars that originate 
through binary mergers or common envelope evolution should show pathological rotation.  
The pulsating sdBV stars, if they represent ``typical'' sdB stars and, by extension, 
typical HB stars, should allow asteroseismology to address the state of internal 
rotation of HB stars.

In fact, one sdBV star, PG~1605 was found to have rapid rotation 
based on the asteroseismic analysis by \citet{Kaw98} -- this was later partially
verified by \citet{hebetal99} through measurement of rotational broadening of spectral
lines.  Armed with this initial success, it is likely that asteroseismology can be
a useful tool to explore important questions 
of the second parameter problem in globular clusters and the origins of sdB stars 
in the field. 

\subsection{This paper}

Most single-star models of HB stars that include rotation were developed for 
Pop II stars. The blue HB stars in globular clusters are quite old, and originate 
from stars that are less than 1$M_{\odot}$.  Given the potential of sdBV stars for 
probing internal rotation, we therefore have computed new models of RGB stars and 
their hot HB decendents that are suited for seismic analysis appropriate to the 
sdBV stars.  

The more recently formed single field HB stars such as the sdB stars probably 
originate from higher mass progenitors (up to the maximum mass that undergoes 
the core helium flash).  This means that the stars that are now sdB stars could 
have had much higher initial angular momenta \citep{Kaw87} as well.  Therefore, 
we have computed models of rotating RGB and sdB stars following the same strategy 
as \citet{SP2000}, but using higher mass Population I progenitors.  Also, we 
compute a dense grid of models to adequately sample the various structures that 
might be relevant for the pulsators.  We then look at the asteroseismic consequences 
of these rotation profiles, and begin to analyze the data on the pulsating sdB stars
to see if they do show rapid core rotation.

This paper proceeds as follows.  In the next section, we describe our stellar
evolution code and our procedures for computing through the RGB and HB. 
We also discuss the limiting cases of angular momentum transport.  
Section 3 outlines the results of these calculations in terms of the surface rotation 
velocities and internal angular velocity profiles of the models.  
Using these profiles, we then use Section 4 to explore the influence of internal
rotation on the observable pulsation frequencies in a preliminary asteroseismological
analysis of the models.  
In Section 5 we discuss the results of the asteroseismological analysis  in terms of the 
observed properties of the pulsators.
We conclude in Section 6 with a discussion of our results, and an outline for more 
detailed study of the pulsators.

\section{Evolutionary Model Calculations}

\subsection{The evolution code and basic model parameters} 

The evolution of our stellar models was computed using ISUEVO, the Iowa State 
University Evolution 
code.  ISUEVO is a standard stellar evolution that is optimized for producing models for 
use in asteroseismology studies.  ISUEVO computes models ranging from the ZAMS through 
the RGB and AGB.  We also compute models of white dwarfs and planetary nebula central 
stars and of course the horizontal branch.  Features of ISUEVO as applied to white 
dwarfs have been described in \citet{Kaw93}, \citet{Kawbra94}, \citet{Dehkaw95} and 
\cite{OBrkaw00}, with further details in \citet{Dehn96}.  ISUEVO models have also 
been used for studies of HB stars \citep{Stoetal97, vanhetal98, Kaw98, Kilk1605, 
ReedetalF48} and AGB stars \citep{Woodetal04, Kawetal03}.

The constitutive physics used in this code are described in the above references, 
but we summarize relevant details here.  Opacities are taken from the OPAL compilation 
for standard mixtures \citep{Igrog96}.  Reaction rates utilize cross sections 
from \citet{Caufow88}.  We follow the standard mixing length treatment with 
$\alpha =1.4$.  During core helium burning on the HB we handle semiconvection 
in the core using a method similar to \citet{Castetal85}.

As we are modeling Population I stars, we assume a metallicity Z=0.02 and an
initial mixture that is solar.  Model sequences were computed with initial masses
of 1.0, 1.5, and 1.8 $M_{\odot}$, and evolved from the ZAMS until degenerate ignition 
of helium in the core (the helium core flash).  The core masses at the helium flash 
for these models are given in Table 1 (measured to the midpoint of the hydrogen--burning 
shell) along with a sample of others from the literature.  For
this paper, we did not compute the changes in the compositional structure caused
by diffusion, and did not include mass loss while on the RGB.
As dictated by observations, in the future we should probably consider models that
deviate from this simple history; for example, in the case of RGB mass loss and late
core helium flash such as computed by \citet{brownetal01}, the core mass of hot 
horizontal branch stars could be
significantly smaller than the standard values.

We did not follow the evolution of the model through the helium flash with our code,
as the helium core flash may be a hydrodynamic event \citep{Deup96}.  Rather, we follow
the conventional technique of computing a Zero Age Horizontal Branch (ZAHB) model that
retains the properties of the RGB core at the flash \citep{Swegro76}.  The
hydrogen shell profile is fit to an analytic function during the construction
of the ZAHB model.  For envelopes thinner than 0.0003$M_{\odot}$, we had to
reduce the thickness of the hydrogen to helium transition zone to preserve the
surface abundance of hydrogen at the pre-flash value.   For most models of interest
for our study of sdB stars, the envelope thickness was too small to support an
active hydrogen burning shell  on the ZAHB, and the stellar luminosity is derived
entirely by the convective helium--burning core.

\subsection{Angular Momentum evolution}

The transport of angular momentum within stars that are evolving up the giant
branch is dominated by the effects of convection in the deep convective envelope.
Starting from some initial state on or just after the main sequence, 
the distribution of specific angular momentum $j$ would stay relatively constant,
affected only by slow diffusive processes (and perhaps magnetic fields if present), 
in the absence of convection \citep{Zahn92}.  

In the radiative regions of our models, we assume that $j$ is conserved locally, which
is roughly  equivalent to the shellular rotation described by \citet{Zahn92}.
We note here that the angular velocity profiles in our models represent the extreme 
case of no diffusion of $j$ and that, in real stars, the angular velocity profiles
will be similar to or shallower than what we report.  The total angular momentum 
within radiative zones is also an upper limit, as diffusive transport of $j$ usually 
drains angular momentum from the core into the more slowly spinning envelope.

Within convection zones, we follow the procedure of SP2K and examine two limiting
cases of $j$ transport.  The first case, corresponding to SP2K's Case B, is 
that convective mixing produces constant angular velocity with radius in convection
zones, with the total angular momentum of the convection zone being conserved.  
There is some observational evidence for this, for example helioseismic
observations of the rotation of the solar convection zone show constant differential
rotation with depth (but not necessarily with latitude) \citep{Thometal96}.
Case B models therefore transport $j$ outwards, as the outer layers of a convection
zone have larger moments of inertia.  For rapidly rotating convective regions (i.e.
those where the rotation velocities are comparable to the speed of convective
eddys) this seems like a reasonable approach.

The second case that we consider is comparable to SP2K, Case D: the 
total angular momentum within the  envelope convection zone is conserved, and $j$ is 
constant within.  In Case D evolution, then, the surface convection zone rotates 
faster in its inner regions where the radius (and therefore moment of inertia) 
is smaller.  In essence, Case D treats $j$ as if it is a compositional element,
mixing thoroughly throughout the convection zone. SP2K point out that such rotation 
in giant star envelopes is consistent with the much slower rotation velocities found 
in the convection zones of these stars.  Even with our larger initial angular 
momenta (see below) the velocities of rotational motion are much smaller than typical 
convective velocities of a few km/s.  This suggests that the small amount of differential 
rotation resulting from the assumption of constant $j$ can be maintained by the 
turbulent mixing generated by convection.

With the treatment of angular momentum described above, and since we do not include
the effects of rotation on the hydrostatic structure on the models, we can treat the
rotational evolution of the models as a ``side calculation'' during model evolution.
In addition, we can compute the angular momentum distribution in a dimensionless way,
by setting the initial model to constant angular velocity of 1.  Subsequent evolution
results in changes in $\Omega$ and $j$ that are proportional to the initial value, 
allowing us to scale our results with a simple multiplicative factor that corresponds
to the initial angular velocity (or, equivalently, the initial angular momentum).

For comparison with observations, we will generally assume that the initial rotation
on the main sequence was as a solid body.  The initial rotation rates for our models
are taken from \citet{Kaw87} -- for example, for the 1.8$M_{\odot}$ model, 
\citet{Kaw87} implies $v_{\rm rot}=180$~km/s, $J_{\rm init}=3.17x10^{50}$, and 
$\Omega=2.49x10^{-5}s^{-1}$.

\subsubsection{Angular mometum transport time scales}

For this preliminary investigation, the only angular momentum transport
we consider is what is required, ad hoc, to produce the two limiting
cases described above.  Other mechanisms must act as well - hydrodynamic
and thermal instabilities that arise would all serve to transport
angular momentum down the angular velocity gradients, effectively reducing
the angular velocity contrasts that develop.  These transport mechanisms
will act on varying time scales.  As long as these time scales are longer
than or comparable to the evolutionary time scale on the RGB and HB, the
real angular velocity profiles will resemble the limiting cases above.
For reference, the shortest Eddington--Sweet circulation time scale within
these models is approximately $10^{11}$ to $10^{12}$ years.  However, given
the steep gradients in $\Omega(r)$ that these models exhibit, other
shear instabilities could act on much shorter time scales.

It is difficult to provide quantitative estimates of these time scales
as they vary within the stellar model, and depend on unknown (and perhaps
unknowable) properties such as the internal magnetic field strengths.
However, for the purposes of this preliminary investigation, we are content
to present and discuss the limiting case scenarios for the angular velocity 
under the two assumptions outlined above, with full acknowledgement that
the real stars will not show such extreme behavior.

\section{Rotation rates and profiles}

\begin{figure}
\plotone{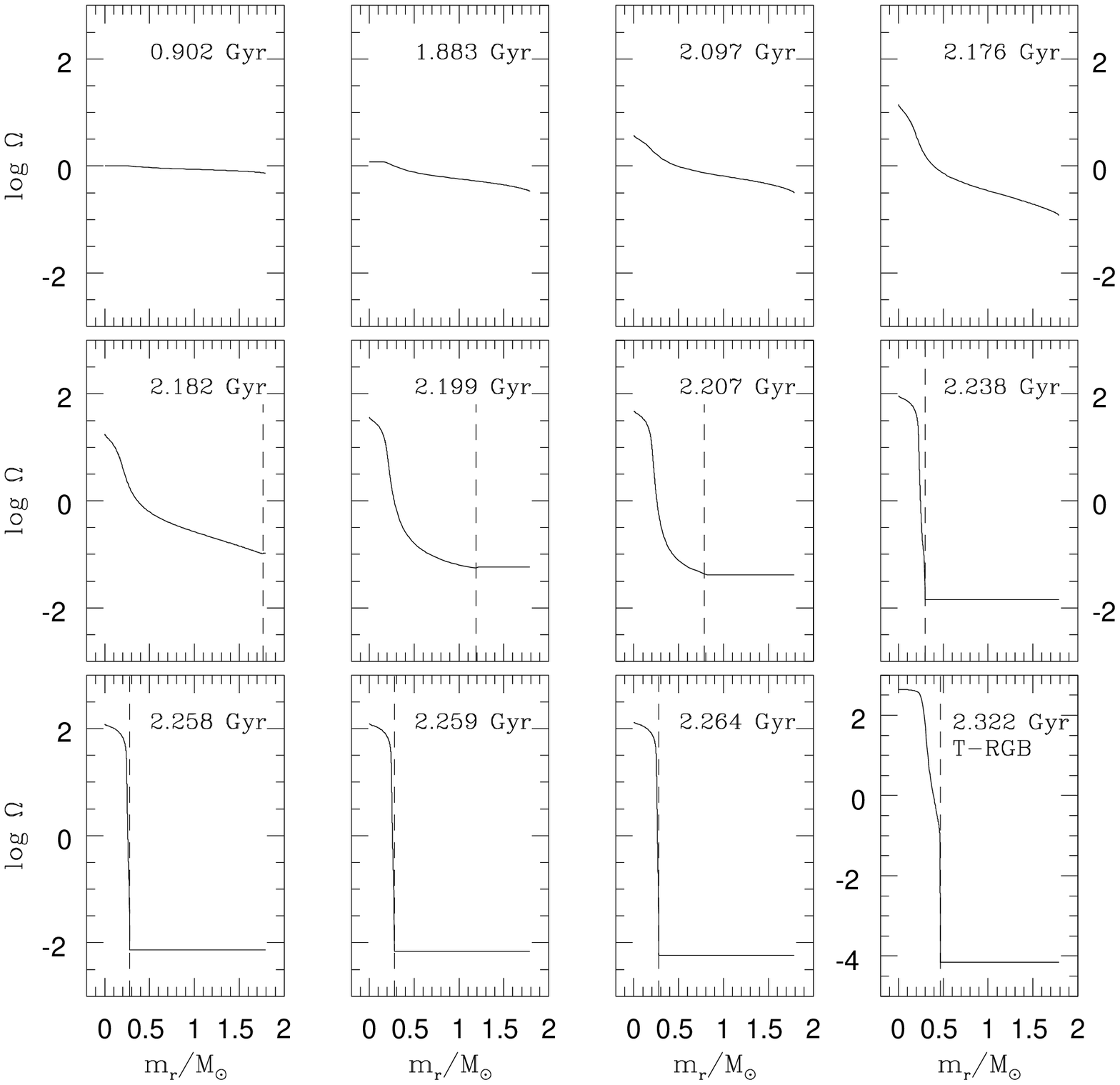}
\caption{Angular velocity as a function of $m_r$ during early RGB evolution
for the $1.8M_{\odot}$ Case B model.  The vertical dashed line denotes the
base of the surface convection zone.  Ages are given in each panel in Gyr.
Note that the establishment of a steep angular velocity gradient occurs quite
quickly as the star reaches the RGB, steepening significantly in a few times
$10^7$ years.  The maximum depth that the convection zone penetrates is in
Panel 10.  The last panel shows the rotation curve of the model immediately 
before the helium core flash. \label{rgbrot}}
\end{figure}

 Generally, our results for RGB rotation profiles are very 
similar to those of SP2K.
In all cases, evolution up the RGB means an increase in radius (and moment of
inertia) meaning that the outer layers slow down.  At the same time, the core
is contracting and therefor spinning up.  Thus evolution up the RGB 
steepens the angular velocity profile.  Figure~\ref{rgbrot} shows the evolution
of the angular velocity during this phase for the 1.8$M_{\odot}$ Case B model 
(constant $\Omega$ in the surface convection zone).  Since most of the moment of 
inertia is in the outermost layers, preserving the bulk angular momentum of the 
outer convection zone at a constant velocity means that the deepening (and 
expanding) star must spin down greatly.  As shown in this figure, the surface 
layers rotate nearly 4 orders of magnitude more slowly than the core, reflecting 
the growth of the star by a factor of nearly 100 in radius. 

\placefigure{rgbrot}

\begin{figure}
\plotone{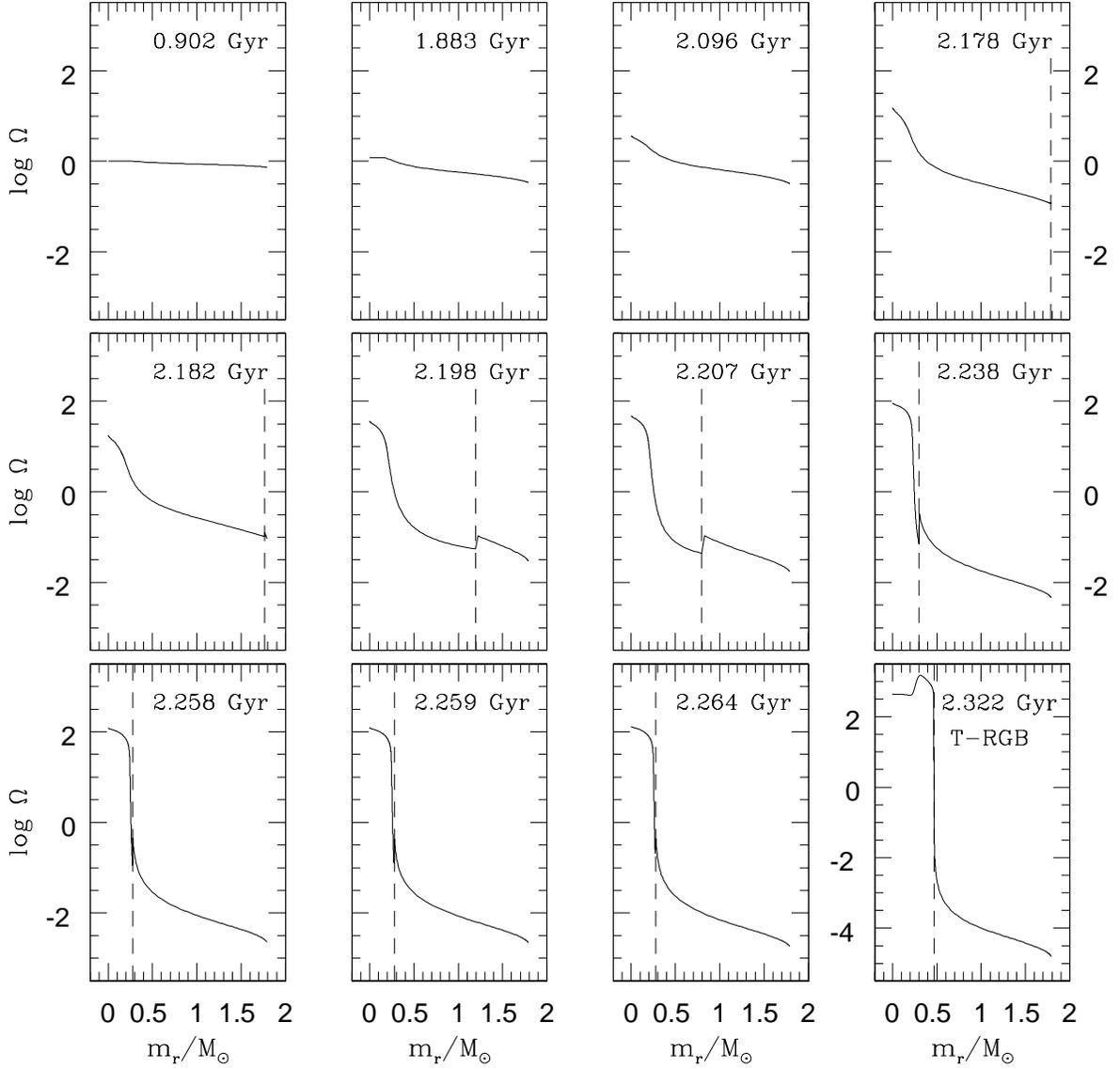}
\caption{Same as Figure~\ref{rgbrot}, but for Case D evolution with constant
specific angular momentum in the outer convection zone. \label{rgbrotjm}}
\end{figure}

The Case D model (constant $j$ through the surface convection zone) exhibits 
very similar behavior overall.  However, the inner
parts of the envelope have greater $j$ than the previous case (Case B), and must 
rotate faster with their smaller moment of inertia.  Figure~\ref{rgbrotjm} shows the
evolution of the angular velocity on the early RGB for the 1.8$M_{\odot}$ Case
D model.  In this case, as the
convection zone deepens, the material that had a smaller $j$ is mixed with the
envelope material with higher specific angular
momentum, raising the angular velocity at the deepening
base.  For
constant specific angular momentum, angular velocity decreases as moment of 
inertia increases, and the specific moment of inertia increases as radius 
squared.  Thus the rotation rate increases with decreasing radius in such 
a convection zone.  Material below the base of the convection zone has slowed
by expansion, and an angular velocity kink develops just below the base of the
envelope convection zone.

\placefigure{rgbrotjm}

As shown by SP2K, the dominant feature of the internal rotation profile 
is the change in angular velocity that occurs near $0.27M_{\odot}$ from the center of 
the model.  The rotation profile drops precipitously at this point 
as the result of draining of angular momentum from the core by the deepening, and 
then retreating, convective envelope.  

During early RGB evolution, the convective envelope appears and deepens.  It
reaches a maximum depth that is modestly dependent on the total stellar mass.
The maxiumum depth that the surface convection zone reaches on the RGB is given
in Table 1.  Clearly, angular momentum transport
in the convective envelope is critical in determining the eventual angular velocity
profile of the model.

\subsection{HB rotation}

\begin{figure}
\plotone{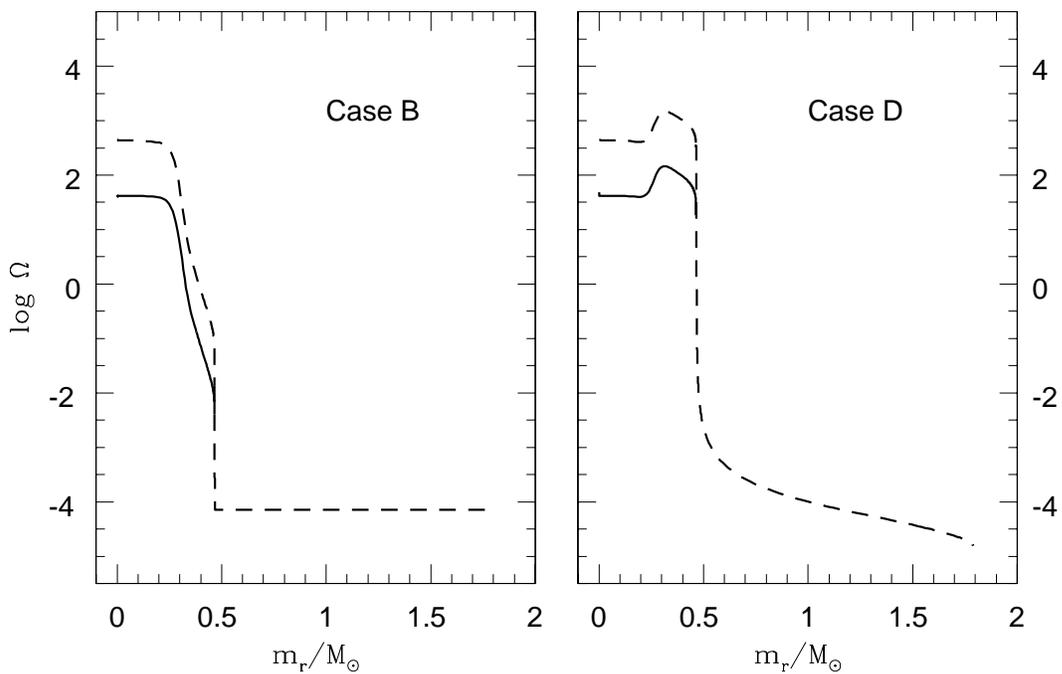}
\caption{Angular velocity as a function of mass before the helium core flash
(i.e. as a RGB core) and after reaching the horizontal branch (as a
helium--core burning model).  The dashed line represents the rotation profile
on the RGB, and the solid line shows the rotation profile after reaching the HB.
The two cases are Case B (left) and Case D (right).
\label{corergbhb}}
\end{figure}

Following the helium core flash, the envelope is ejected, and the remnant core
settles onto the horizontal branch.  We have assumed that the specific
angular momentum is preserved everywhere in the remnant core, and thus the
angular velocity profile will reflect changes in the structure of the core as
it converts from a degenerate configuration to a nondegenerate, helium burning
region.  As a degenerate core on the RGB, the structure resembles a polytrope
with the same polytropic index as the convective helium burning core on the
EHB (i.e. \citet{hankaw}).  Therefore, the radius profile (as a function of
mass) is a power law of the same power but different coefficient.  The
nondegenerate helium--burning core will have a larger radius, and therefore
rotate more slowly (by the same factor) in the core than it did just prior to
the helium core flash. Figure~\ref{corergbhb} shows this for the two cases of
angular momentum transport in convection zones.
\placefigure{corergbhb}

\subsubsection{Case B HB models}

Horizontal branch stars evolved from Case B RGB models show much smaller 
surface rotation velocities than those with Case D precursors.  These models
preserve a rapidly rotating core with roughly constant angular velocity out
to about 0.3$M_{\odot}$.  Material outside of that point had been drained of
angular momentum by inclusion in the convective envelope early during RGB
evolution, and so remain slowly rotating on the HB.  With such small exterior 
rotation velocities, these stars contain a much larger contrast between the 
surface and core rotation velocities.  Angular velocity drops by almost a factor 
of $10^5$, as shown in Figure~\ref{omvsrhb}.

\subsubsection{Case D HB models}

The rotation profiles of models descended from Case D RGB evolution contain
relics of the high specific angular momentum of material in the outer layers
of the RGB progenitor.  

When the constant J/M envelope reached down to the
0.3$M_{\odot}$ zone, and then retreated, the higher specific angular momentum
left behind produced a local maximum of angular velocity that persists into
EHB evolution.  The
inner core never had a chance to share in this high angular momentum pool,
and was left rotating more slowly.  

On the EHB, the core becomes convective, but does not reach the
maximum angular velocity material.  Therefore, the angular velocity profile
changes little during EHB evolution - the core does spin down a bit as the
angular momentum is mixed, but the inner core continues to spin about a
factor of 5 slower than the fastest material.  The angular velocity drops
from the maximum by a factor of about 10 or so to the surface.

\subsection{Evolution of the angular velocity profile on the EHB}

\begin{figure}
\plotone{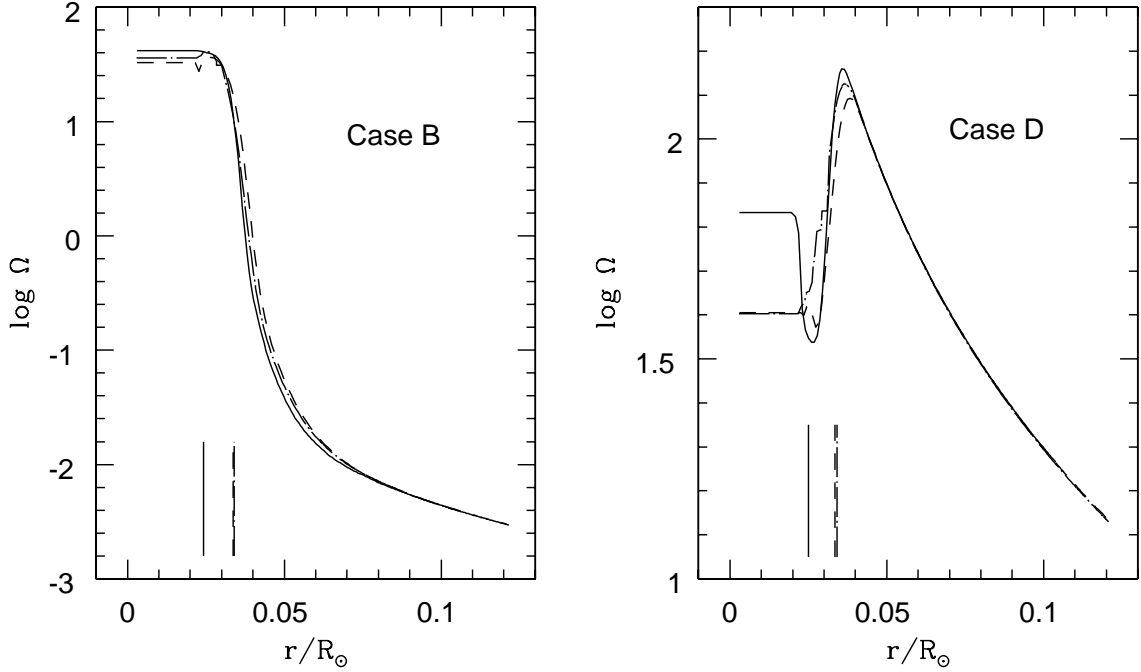}
\caption{ Rotation profiles for EHB models, for two cases for RGB angular
momentum evolution and at different HB evolutionary stages.  The left panel
corresponds to Case B evolution and the right panel to Case D evolution.
Models at three stages of evolution are shown, with the short vertical
lines at the bottom corresponding to the transition from pure helium to the
helium--depleted core.   Note that the profiles show very subtle changes
in the outer layers with evolution.  For Case D evolution, the inner core
shows changes reflecting mixing of angular momentum by the convective core
as it grows and then shrinks.  In Case B RGB evolution, the core spins 4
orders of magnitude faster than surface.  Case D RGB evolution results in
only a factor of 10 decrease in the rotation velocity from the interior
to the surface of the model. \label{omvsrhb}} 
\end{figure}

The general features of the internal rotation profile remain nearly fixed
through the stage of core helium burning.  As shown in Figure~\ref{omvsrhb},
the rotation profile (with radius) shows the same general features at the
start and end of HB evolution.  For Case B evolution, a rapidly rotating core
is preserved.  For Case D evolution, the outer decrease in angular velocity
is preserved, while the inner, more slowly rotating core, corresponding to
the fully convective helium--burning zone, spins down at the start of EHB
evolution as the core grows and mixes in lower angular momentum material.
Further evolution results in little additional angular velocity change with
radius.  Given the uncertainty about the initial conditions (e.g. initial 
angular momentum on the main sequence, and main sequence angular momentum 
evolution), this relatively small change in the angular velocity profile on 
the EHB could allow efficient parameterization for studies of their seismic 
influence.

\placefigure{omvsrhb}

Given that our plan is to use these models for asteroseismology, the structure 
of the internal rotation profile is the focus of our study rather than the surface 
rotation velocities.  Still, it is useful to examine the surface rotation rates
of these models for general comparison with observation.

Given the way we construct HB models, the surface rotation velocity of models
of different mass reflect the value of $j$ at the surface -- which in turn
results from the truncation of the envelope of the RGB progenitor.  For
Case B evolution, $j$ decreases with depth in the RGB progenitor.  HB models
with thicker envelopes therefore have surface layers with larger $j$, and 
faster rotation.  However, since we are interested here in the bluest HB stars,
the envelope masses are all less than 0.005$M_{\odot}$.  The change in
$j$ in the seed model over that small range is less than a factor of 2.  In the
Case D models, $j$ is constant outside of the core - reflecting the assumptions
built into the RGB model - so all of our Case D models have the same value
of $j$ in their surface layers.

\begin{figure}
\plotone{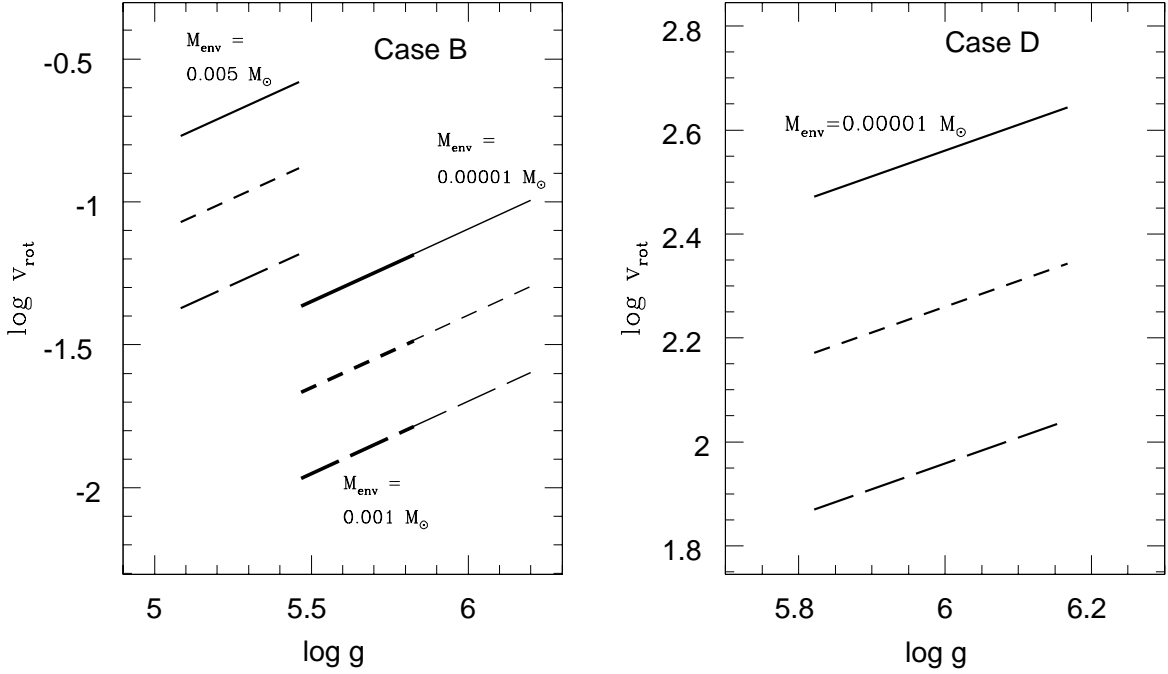}
\caption{Surface rotation velocity as a function of $\log g$ for sdB models.
The left panel shows Case B models derived from the 1.8$M_{\odot}$ sequence
with surface masses as indicated, and the right panel shows representative
Case D models from the same mass seed model.  The solid line shows models
with initial (main sequence) rotation velocities of 200 km/s; the short 
dashed lines had initial rotation velocities of 100 km/s, and the long 
dashed lines had initial rotation velocities of 50 km/s. \label{logvlogg}} 
\end{figure}

\placefigure{logvlogg}

The surface angular velocity of the HB models depends on the initial
value of $j$ at the surface (since $j \propto R^2 \Omega$).  The surface
rotation velocity $v_{\rm rot}$ scales as $R^{-1} \Omega$.  Spectroscopic
measurements of the sdB stars can determine $\log g (\propto R^{-2})$ to
adequate precision to compare with our models.  Therefore, for a fixed $j$,
the rotation velocity should scale with $g^{-1/2}$ with the constant of
proportionality depending on the models and the initial angular momentum.
The rotation velocities for sample models are shown in Figure \ref{logvlogg}
for a variety of initial rotation velocities.

\section{Asteroseismology Probes}

\begin{figure}
\epsscale{0.8}
\plotone{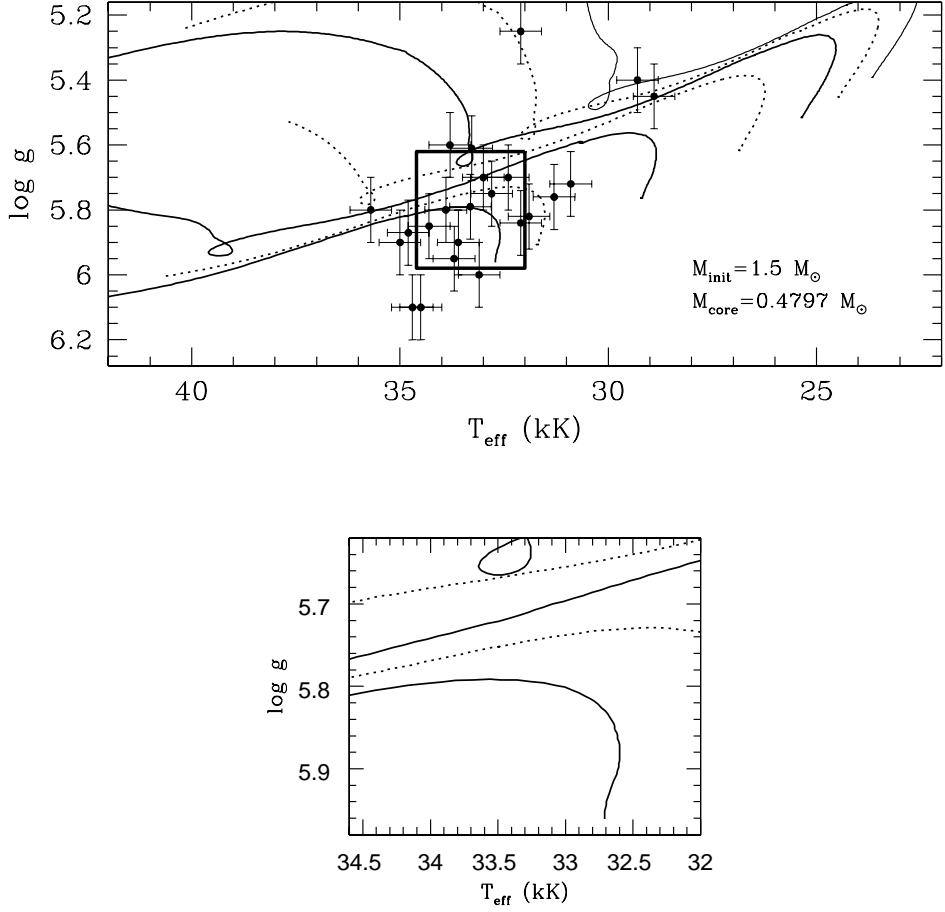}
\caption{Evolutionary tracks of EHB models representative of the subdwarf
B stars.  Models include a core derived from the post-helium flash core
of a 1.5 $M_{\odot}$ star with increasing hydrogen-rich envelopes with
increasing $T_{\rm eff}$.   Alternating tracks can be distinguished by
the different line types.  Envelope masses are 0.00001, 0.00004, 0.00032,
0.00122, 0.00232, 0.00322, 0.00422 $M_{\odot}$, running from lower left
to upper right.  Data points are the pulsating sdB stars (from Kilkenny
2001), The box is centered at the mean position of the sdB pulsators,
and the size of dimensions of the box are proportional to the width of
the distributions.  The lower plot is an enlargement of the box, showing
that sdB models of various evolutionary stages underly the positions of
the pulsating sdB stars.  \label{hrdsdbv}} 
\end{figure}

The pulsating sdB stars - known as the EC~14026 stars or simply sdBV stars,
are found in the midst of the ($\log g$,$T_{\rm eff}$) plane locus of the
``ordinary'' sdB stars.  They range in $T_{\rm eff}$ of slightly less than
30,000K to over 35,000K, and have $\log g$ between 6.1 and 5.3. Most lie at
the high-gravity end, clustering near the ZAEHB, with the cooler and lower
gravity members numbering about 10\% of the total.  Figure~\ref{hrdsdbv}
shows an H--R diagram that includes a representative sample of EC~14026
stars and sample
evolutionary tracks from our sets of models.  Given the convergence of
the tracks in this plane, it is clear that H--R diagram position alone is
insufficient to judge their evolutionary state, with stars that are close
to the ZAEHB intermixed with stars that have depleted helium in their cores.

\placefigure{hrdsdbv}

The lowest gravity EC~14026 stars fall sufficiently far from the ZAEHB that
they cannot be core helium--burners for core masses that are consistent with
single--star evolution.  Thus they are either on their way to the AGB or
will evolve directly to the white dwarf cooling track in a very short time.

Another class of multiperiodic pulsating sdB stars exhibit longer periods
(between 0.3 and 2.5 hours).  These stars, colloquially known as the
``Betsy stars'' after their discoverer \citep{Betsy03}, lie adjacent to the
EC~14026 stars in the H--R diagram at the cool end.  Fontaine et al. (2003)
postulate that these are $g$-mode pulsators driven by the same mechanism
that drives the pulsations in the EC~14026 stars.

The pulsators that lie in the main clump are shown in Figure~\ref{hrdsdbv} 
surrounded by a box that represents the mean values of $T_{\rm eff}$
and $\log g$; the size of the box represents the standard deviation of
these means.  To explore the influence of a rapidly rotating core on the
pulsations of sdBV stars, we employ representative models that fall within
this box.

Many of the high--gravity pulsating sdB stars show complex pulsations that
require a large number of independent periodicities.  Examples include
PG~1047 \citep{Kilk1047} and PG~0014 \citep{Brass0014}.  The rich pulsation
spectra show a mode density that is too high to be explained using only
radial pulsations.  Nonradially pulsating stars have a large number of
available pulsation modes -- manifested as a large number of possible
oscillation frequencies.  Even normal nonradial modes have a limited
frequency distribution, meaning that some of the pulsators cannot be
understood without resorting to either high values of $\ell$  (a large
number of nodal lines on the stellar surface) or to rapid rotation.

Generally, for nonradial pulsations, each mode samples a slightly different 
part of the stellar interior, and so we can in principle use the observed
pulsations to probe the structure of the stellar interior.  Of particular
relevance to this paper, nonradial modes are sensitive to stellar rotation -
rotation can significantly increase the number of distinct oscillation
frequencies seen in pulsating stars by lifting the degenercy of modes
with the same degree ($\ell$) but different order ($m$) as discussed below.

In this section, we describe how nonradial pulsations in sdB stars may
provide a window into their interiors, and test the hypothesis that
horizontal branch stars have rapidly rotating interiors.  The discussion here
is intended to be somewhat general so as to illustrate the process by which
we can detect rapid rotation of the stellar interior using currently observed
sdBV stars.  A future paper will describe our asteroseismic results in more
detail and apply the analysis to individual pulsators.

\subsection{Rotational Splitting of Nonradial modes}

Nonradial pulsations in stars are generally expressed in terms of spheroidal
modes.  The spectrum of available oscillations are characterised by the
eigenfrequency $\sigma_{n \ell m}$ corresponding to a spheroidal mode with
quantum numbers $n$, $\ell$, and $m$.  The values $\ell$ and $m$ refer to the
spherical harmonic $Y^m_{\ell}$ with $\ell$ nodal lines, $m$ of which pass
through the axis of symmetry. The sign of $m$ gives the direction of
propogation of the corresponding surface running wave, and therefore there
are $2\ell+1$ possible values of $m$ for a given $\ell$.  The radial order of
the mode (corresponding roughly to the number of nodes in the radial
direction) is $n$. 

For perfect spherical symmetry, the oscillation frequencies are independent
of $m$ and depend only on $n$ and $\ell$.  However, if rotation (or another
non--spherical process) is present, the value of the oscillation frequency
will depend on $m$.  In the case of slow rotation as a solid body, the
$m$ dependence of the oscillation frequency can be written as:
\begin{equation}
\sigma_{n \ell m} = \sigma_{n \ell 0} + m \Omega (1-C_{n \ell})
\end{equation}
where the quantity $C_{n,\ell}$ is a function of the spatial structure of the
oscillations within the star (the eigenfunctions).  For the above case of
slow rotation,
\begin{equation}
C_{n \ell} =\frac{\int_0^R{(2 \xi_r \xi_h + \xi_h^2) \rho r^2 dr}}
                 {\int_0^R{(\xi_r^2 + \ell [\ell +1] \xi_h^2) \rho r^2 dr}}
\end{equation}
where the radial and horizontal perturbations at $r$ are given by
$\xi_r$ and $\xi_h$ respectively.  For a more complete description of
rotational splitting, see one of the standard texts of stellar pulsation 
(i.e. \citet{Cox80,Unetal89}).

Equations (1) and (2) correspond to solid body rotation with a rotation
frequency that is small compared to the oscillation frequency.  The constancy
of $\Omega$ means that it does not appear inside the integral in Equation
(2), allowing simple determination of the effect of rotation in this special
case.  Clearly, rotation can split an $\ell=1$ oscillation into a triplet,
spaced in frequency by the rotation frequency times $(1-C_{n \ell})$.
Sequences of such equally spaced triplets have been clearly identified 
in a number of pulsating white dwarf stars, for example, suggesting that they
rotate nearly as solid bodies, at least in those portions of those stars that
contribute to the integral (2) \citep{KSG99}.

If a star undergoes differential rotation with radius -- that is, 
if $\Omega = \Omega (r)$, then the simple form for calculating the value of 
$\sigma_{n \ell m}$ above is not appropriate.  The splitting in this case 
(where $\Omega$ depends on radius, not on latitude) is given by
\begin{equation}
\sigma_{n \ell m} = \sigma_{n \ell 0} 
             + m \int_0^R{\Omega(r) K_{n \ell}(r) dr}
\end{equation}
where
\begin{equation}
K_{n \ell} = \frac{(\xi_r^2 
                  -[\ell (\ell +1)\ -1] \xi_h^2 - 2 \xi_r \xi_h) \rho r^2}
                 {\int_0^R{(\xi_r^2 + \ell [\ell +1] \xi_h^2) \rho r^2 dr}}
\end{equation}
is the rotation kernel corresponding to a mode with  $n$ and $\ell$. 

Thus for a star with internal differential rotation with
radius, the observed splitting
for a mode with a given $n$ and $\ell$ represents an average of $\Omega(r)$ 
weighted by the rotation kernel $K_{n \ell}(r)$.  If the kernel function of a
mode is nonzero where the star has very rapid rotation, then the resulting
splitting can reveal the inner rapid rotation despite slow rotation at the 
surface layers.

\subsection{Sample rotation kernels and results for sdB models}

\begin{figure}
%\epsscale{0.8}
%\includegraphics[angle=270, width=17cm]{f7.eps}
\plotone{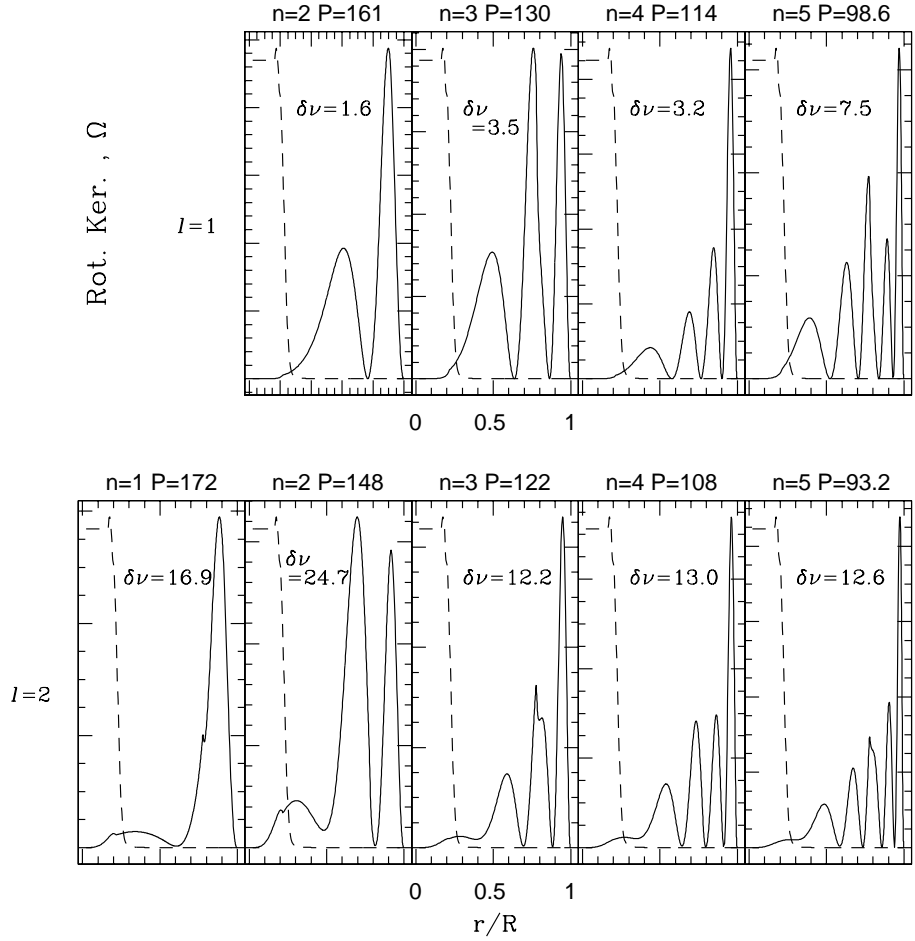}
\caption{Ssample rotation kernels for a representative nearing the end 
of core helium burning, for modes with periods and degrees as indicated.  
Dashed lines represent $\log \Omega(r)$ for Case B angular 
momentum transport in the RGB convection zone. Splitting values are in 
$\mu$Hz. \label{casebmkern}} 
\end{figure}

Figure~\ref{casebmkern} shows sample rotation kernels for an sdB model
representative of the pulsating EC 14026 stars.  This particular model
is nearing the end of core helium burning as a middle--aged sdBV model.
The modes illustrated span the period range seen in the pulsating sdBV stars.
Note that higher $n$ modes have more peaks, but that in all modes the
kernel has nonzero value close to the stellar core.  Also indicated in
Figure~\ref{casebmkern} are lines representing $\log \Omega(r)$ for Case
B angular momentum transport in the RGB outer convection
zone.  Note that even though the amplitude of $K_{n \ell}$ is small in
the inner regions, the large value of $\log \Omega(r)$ in the same region
suggests that the value of the rotational splitting will be dominated by
the core rotation.

The case for $\ell=2$ modes is more striking.  For several modes in this
model, the rotation kernel is quite large in the core.  This likely results
from resonant mode trapping by the composition transition zones, in
particular the He/C+O zone in the core.  At this evolutionary stage, the
model shows some mixed--character modes for $\ell=2$ which can have
significant amplitude in the core \citep{CharpII}.

Tables 2-7 present the computed splittings for models of different evolutionary
stages that lie within the box of Figure \ref{hrdsdbv}.  The first two tables
give periods and splittings for a near-ZAHB model; Tables 4 and 5 show the
same range of periods in a middle-aged sdB model, and Tables 5 and 6 show results
for a highly evolved sdB model.  Tables show the splittings expected if
the model is rotating as a solid body at the surface rate, and for models with
differential rotation.  The splittings
for differential rotation in models with Case B are given in Tables 2, 4, and 6;
Case D models are in Tables 3, 5, and 7.

As an illustration of the kind of effects that differential rotation can produce, 
consider Tables 4 and 5, which present the computed splittings for two models of 
middle-aged sdB stars.  In the Case B model in Table 4, the surface
rotation velocity, if representative of the entire star, would produce
a splitting of less than $0.01 \mu{\rm Hz}$ for all modes (surface rotation
velocity of only 6 m/s), yet the rapid internal rotation results
in splittings ranging from 1 to 80 $\mu{\rm Hz}$ depending on the value of
$n$ and $\ell$ for the mode. 

Table 5 shows what happens for a Case D model, which has an artificially
scaled-down rotation rate of 10 km/s at the surface (this model would have
been rotating way too quickly if it had the initial angular momentum assigned
from the main sequence progenitor).  For this model, the 
expected splitting would be about 15 $\mu{\rm Hz}$ given the surface rate
and the assumption of solid body rotation.  The rapidly rotating core,
however, produces splittings ranging from 24 to 52 $\mu{\rm Hz}$.

In both limits, the rotational splitting is much larger than
expected from assuming solid body rotation at the surface rotation 
velocity.  Even more interesting
is the fact that the splitting changes (sometimes dramatically) from
one mode to the next within a sequence of the same $\ell$ but different
$n$.  This reflects the fact that the low--order modes in these stars
sample portions of the core and envelope, and the exact parts sampled
depend on the overtone number.  With a steep drop in $\Omega$, the
kernel functions (see Figure \ref{casebmkern} as an example) sweep
across these fluctuations as $n$ increases.  In addition, mode trapping
by the composition zones present in these models produces some
strong core localization of the kernels for certain modes, and these
therefore show larger rotational splitting.

The fact that the expected splittings can be quite large, coupled 
with the large differences in expected splittings from mode to mode, 
can make observational identification of rotational splitting quite 
difficult.
Generally, the way this has been done in the past is to look for
a sequence of equally--spaced triplets in a Fourier transform, and
then identify them as $\ell=1$ for example.  Even if all members 
of such triplets are not seen, finding several pairs of modes split
by the same amount (or twice that amount, or half) would be 
indicative of rotational splitting.  For sdBV stars, though,
with significant mode--to--mode differences in splittings expected,
a series of rotationally split frequencies would be virtually 
indistinguishable from modes of different $\ell$ and $n$. 
For example, consider the model of an advanced-stage sdB star whose
periods are listed in Tables 6 and 7.  The $\ell=2$ modes have 
periods that are close to $\ell=0$ (radial) modes.  With splittings
of 40 to 200 $\mu$Hz for the lowest order modes, the quintuplets
$\ell=2, n=0, 2$ and 3 could spread over very large frequency ranges
that would encompass, or nearly reach, $\ell=0, n=1$; $\ell=0, n=2$; 
and $\ell=1, n=3$. 
Without appealing to differential rotation, such a rich mode
spectrum would require appealing to other mechanisms to explain
the periodicities.

\section{Conclusions}

We have shown that sdB stars, like horizontal branch stars in
general, should retain rapidly rotating cores as a relic of their
evolution on the RGB.  Such rapid rotation could serve as a
reservoir of angular momentum which, when tapped, can produce
anomalously fast rotation at the surface on the HB.  Taken alone, 
our results are consistent with the models of SP2K in that without
late angular momentum transport, standard models of single
HB stars cannot explain observed trends in the distribution of 
surface rotation velocities with effective temperature.  However,
if angular momentum transport can tap the internal reservoir in
the absence of composition gradients, the suggestion of SP2K that
redder HB stars can spin up in the latter stages of core helium
burning remains theoretically viable.

Other mechanisms can be at work that would serve to reduce or
eliminate the strong differential rotation present in these 
limiting--case
models.  Also, other scenarios for the origin of sdB stars (such
as binary mergers) and of sdB surface rotation velocities (i.e.
spin--up by nearby planets) make specific predictions about the
internal angular velocity profiles.
In any case, though, we can test the predictions of a wide
variety of stellar models if we can measure the rotation rate of the
subsurface layers of HB stars and, in particular, the hot subdwarfs.

Pulsations can probe the interiors of these stars and thereby reveal
the rotation velocity profile within.
Rapidly rotating cores produce rotational splitting of nonradial
modes, and the amount of the splittings is not what would be
expected from spectroscopic measurements of $v \sin i$.
  Evidence that we see such
effects could be the rotation of PG~1605 as deduced from the pulsations.
The rotatonal splitting identified in that star by \cite{Kaw98} is
about 3 times larger than the $v \sin i$ measured via spectroscopy
\citep{hebetal99}.  While this could simply be the result of a small
inclination angle between the rotation axis of the star and our line
of sight, it could also result from differential rotation and a rapidly
rotating core.

Clearly, it is important to consider the possibly large rotational 
splittings that might appear in the pulsation spectra of the EC~14026 
stars.  Identification of modes with large splittings, or of sequences 
of modes with varying splitting would 
point towards strong differential rotation. In either case, such a 
measurement would confirm the existence of a rapidly rotating core 
within these stars. Failure to find such effects would be convincing 
evidence that angular momentum transport mechanisms on the RGB or 
during the HB phase are very efficient at wiping out angular velocity 
gradients. Should that be the case, then the mystery of the rotation 
of horizontal branch stars will remain with us.

\begin{acknowledgements}
This work is partially supported by NSF Grant AST20205983
and by the NASA Astrophysics Theory Program through 
grant NAG-58352, both to Iowa State University.
\end{acknowledgements}

\clearpage

%%%  FIGURES  %%%

%Figure - multipanel omega vs. M on early RGB - constant W

%Figure - multipanel omega vs. M on early RGB - constant J/M

%FIGURE - log omega vs. log m on the RGB and the ZAEHB

% omvsrhb - rotation rate with radius on HB - little change

%Figure - log vrot vs. log go

%Figure - HB Evol tracks and the pulsators

%Figure - l=1 kernels, etc.

%%%%  TABLES  %%%%%
%Table 1 - Model Parameters
\begin{deluxetable}{ccccc}
\tablecaption{Parameters of the evolutionary models with Z=0.02}
\tablewidth{0pt}
\tablehead{
 & & \multicolumn{3}{c}{$M_{\rm c, flash}/M_{\odot}$}\\
\colhead{Initial Mass} & 
\colhead{$M_{\rm CE}^{\rm min}/M_{\odot}$} &
\colhead{this work} &
\colhead{Bono et al. 1997} &
\colhead{Yi et al. 2003}\\
 & & X$_{\rm ZAMS}$=0.74 &
     X$_{\rm ZAMS}$=0.69 &
     X$_{\rm ZAMS}$=0.71 
}
\startdata
1.0 $M_{\odot}$  & 0.248 & 0.4832 & 0.478 & 0.469 \\
1.5 $M_{\odot}$  & 0.269 & 0.4798 & 0.476 & 0.462 \\
1.8 $M_{\odot}$  & 0.278 & 0.4652 & 0.469 & 0.440 \\
\enddata
\end{deluxetable}

%Table 2 - Young Models - Case B (const W in CZ)
\begin{deluxetable}{rrcccc}
\tabletypesize{\small}
\tablecaption{Representative near-ZAEHB model\tablenotemark{a}
 pulsations, Case B}
\tablewidth{0pt}

\tablehead{
\colhead{Period} & \colhead{$\nu$ [$\mu$Hz]} & \colhead{$\ell$}  & 
\colhead{$n$} & \colhead{$1/P_{\rm surf} (1-C_{n,\ell})$}
&\colhead{$\Delta \nu$ [$\mu$Hz]}
}

\startdata
86.45     & 11567  &  0 & 4 &  --  & --   \\
87.30     & 11457  &  2 & 4 & 0.086 & 17.6 \\
91.24     & 10963  &  1 & 4 & 0.087 & 7.8 \\ 
97.25     & 10283  &  0 & 3 &  --  &  --  \\
98.48     & 10155  &  2 & 3 & 0.084 & 19.8 \\
105.5     &  9480  &  1 & 3 & 0.086 & 8.6 \\ 
118.6     &  8434  &  0 & 2 &  --  & --   \\
121.3     &  8246  &  2 & 2 & 0.081 & 25.9 \\
130.1     &  7685  &  1 & 2 & 0.087 & 2.8 \\ 
136.9     &  7302  &  0 & 1 &  --  & --   \\
138.2     &  7237  &  2 & 1 & 0.084 & 11.2 \\
144.3     &  6930  &  1 & 1 & 0.087 & 2.1 \\
158.5     &  6309  &  0 & 0 &  --  & --   \\
168.6     &  5930  &  2 & 0 & 0.059 & 91.4 \\

\enddata
\tablenotetext{a}{
$M=0.46502 M_{\odot}$, $M_H=0.00002 M_{\odot}$\\
$\log g = 5.92$, $T_{\rm eff}$ = 32,025 K \\
$v_{\rm rot}$ = 0.048 km/s,
$1/P_{\rm surf}$ = 0.089 $\mu$Hz
}
\end{deluxetable}

%Table 3: Young models - const J/M in CZ

\begin{deluxetable}{rrcccc}
\tabletypesize{\small}
\tablecaption{Representative near-ZAEHB model\tablenotemark{b} 
pulsations, Case D.}
\tablewidth{0pt}

\tablehead{
\colhead{Period} & \colhead{$\nu$ [$\mu$Hz]} & \colhead{$\ell$}  & 
\colhead{$n$} & \colhead{$1/P_{\rm surf} (1-C_{n,\ell})$} & 
\colhead{$\Delta \nu$ [$\mu$ Hz]}
}

\startdata
84.73     & 11800  &  0 & 4 &  --  & --   \\
85.50     & 11699  &  2 & 4 & 18.0 & 29.3 \\
89.30     & 11198  &  1 & 4 & 18.1 & 32.2 \\
95.88     & 10430  &  0 & 3 &  --  & --   \\
97.22     & 10287  &  2 & 3 & 17.6 & 32.0 \\
104.5     &  9574  &  1 & 3 & 17.9 & 39.8 \\ 
117.1     &  8540  &  0 & 2 &  --  & --   \\
119.5     &  8372  &  2 & 2 & 17.1 & 30.2 \\
126.9     &  7880  &  1 & 2 & 18.1 & 28.9 \\
133.3     &  7504  &  0 & 1 &  --  & --   \\
134.9     &  7414  &  2 & 1 & 17.4 & 25.0 \\
142.4     &  7023  &  1 & 1 & 18.1 & 34.8 \\
157.6     &  6344  &  0 & 0 &  --  & --   \\
167.9     &  5956  &  2 & 0 & 12.3 & 47.3 \\
\enddata
\tablenotetext{b}
{$M=0.46413 M_{\odot}, M_H=0.00002 M_{\odot}$\\
$\log g = 5.93$, $T_{\rm eff}$ = 32,114 K\\
$v_{\rm rot}= 10$ km/s,
$1/P_{\rm surf} = 18.6$ $\mu$Hz,
$J/J_{\rm init}=0.045$
}
\end{deluxetable}

%Table 4 - middle-age, case B

\begin{deluxetable}{rrcccc}
\tabletypesize{\small}
\tablecaption{Representative late-ZAEHB model\tablenotemark{c}
 pulsations, Case B}
\tablewidth{0pt}

\tablehead{
\colhead{Period} & \colhead{$\nu$ [$\mu$Hz]} & \colhead{$\ell$}  & 
\colhead{$n$} & \colhead{$1/P_{\rm surf} (1-C_{n,\ell})$}
&\colhead{$\Delta \nu$ [$\mu$Hz]}
}
\startdata
82.37     & 12140 &  0 & 6 &  --  & --   \\
84.20     & 11876 &  2 & 6 & 0.051 & 13.6 \\
87.85     & 11384 &  1 & 6 & 0.052 & 7.1  \\
91.14     & 10972 &  0 & 5 &  --  & --   \\
93.21     & 10728 &  2 & 5 & 0.051 & 12.6 \\
98.65     & 10138 &  1 & 5 & 0.052 & 7.5  \\
105.6     &  9467 &  0 & 4 &  --  & --   \\
108.3     &  9238 &  2 & 4 & 0.050 & 13.0 \\
113.9     &  8784 &  1 & 4 & 0.052 & 3.2  \\
118.8     &  8417 &  0 & 3 &  --  & --   \\
121.9     &  8206 &  2 & 3 & 0.050 & 12.2 \\
130.3     &  7677 &  1 & 3 & 0.052 & 3.5  \\
141.8     &  7052 &  0 & 2 &  --  & --   \\
147.6     &  6774 &  2 & 2 & 0.046 & 24.7 \\
161.4     &  6196 &  1 & 2 & 0.052 & 1.6  \\
167.2     &  5980 &  0 & 1 &  --  & --   \\
171.5     &  5832 &  2 & 1 & 0.049 & 16.9 \\
179.4     &  5573 &  1 & 1 & 0.052 & 0.9  \\
185.9     &  5379 &  0 & 0 &  --  & --   \\
191.1     &  5232 &  2 & 0 & 0.044 & 75.9 \\
\enddata
\tablenotetext{c}{
$M=0.47991 M_{\odot}$, $M_H=0.00001 M_{\odot}$\\
$\log g = 5.80$, $T_{\rm eff}$ = 33,113 K \\
$v_{\rm rot}$ = 0.034 km/s, 
$1/P_{\rm surf}$ = 0.0532 $\mu$Hz
}
\end{deluxetable}

%Table 5 - middle-aged, case D

\begin{deluxetable}{rrcccc}
\tabletypesize{\small}
\tablecaption{Representative late-ZAEHB model\tablenotemark{d}
 pulsations, Case D}
\tablewidth{0pt}

\tablehead{
\colhead{Period} & \colhead{$\nu$ [$\mu$Hz]} & \colhead{$\ell$}  & 
\colhead{$n$} & \colhead{$1/P_{\rm surf} (1-C_{n,\ell})$}
&\colhead{$\Delta \nu$ [$\mu$Hz]}
}
\startdata
82.65     & 12099  &  0 & 6 & -- & -- \\
83.33     & 11999  &  2 & 6 & 15.3 & 31.0 \\
86.89     & 11508  &  1 & 6 & 15.5 & 31.5 \\
91.45     & 12099  &  0 & 5 & -- & -- \\
92.26     & 10839  &  2 & 5 & 15.3 & 30.2\\
97.68     & 10239  &  1 & 5 & 15.4 & 38.7 \\
106.0     &  9435  &  0 & 4 & -- & -- \\
107.2     &  9331  &  2 & 4 & 15.1 & 29.1 \\
112.6     &  8881  &  1 & 4 & 15.5 & 29.5 \\
119.1     &  8390  &  0 & 3 & -- & -- \\
120.6     &  8287  &  2 & 3 & 14.9 & 27.3 \\
129.0     &  7753  &  1 & 3 & 15.5 & 36.5 \\
142.3     &  7026  &  0 & 2 & -- & -- \\
146.5     &  6828  &  2 & 2 & 13.9 & 31.4 \\
159.7     &  6261  &  1 & 2 & 15.5 & 30.4 \\
167.8     &  5961  &  0 & 1 & -- & -- \\
169.8     &  5890  &  2 & 1 & 14.6 & 23.6 \\
177.5     &  5633  &  1 & 1 & 15.5 & 27.3 \\
186.6     &  5360  &  0 & 0 & -- & -- \\
190.7     &  5243  &  2 & 0 & 12.7 & 52.0 \\
\enddata
\tablenotetext{d}{
$M=0.47979 M_{\odot}$, $M_H=0.00001 M_{\odot}$\\
$\log g = 5.81$, $T_{\rm eff}$ = 32,945 K \\
$v_{\rm rot}$ = 10 km/s, 
$1/P_{\rm surf}$ = 15.8 $\mu$Hz, $J/J_{\rm init}=0.067$ 
}
\end{deluxetable}

%Table 6 - Old model, Case B

\begin{deluxetable}{rrcccc}
\tabletypesize{\small} 
\tablecaption{Representative post-ZAEHB model\tablenotemark{e} pulsations, Case B}
\tablewidth{0pt}

\tablehead{
\colhead{Period} & \colhead{$\nu$ [$\mu$Hz]} & \colhead{$\ell$}  & 
\colhead{$n$} & \colhead{$1/P_{\rm surf} (1-C_{n,\ell})$}
&\colhead{$\Delta \nu$ [$\mu$Hz]}
}
\startdata
83.49     & 11977  &  0 & 7 &  --  & --   \\
83.75     & 11934  &  2 & 7 & 0.042 & 9.3  \\
87.29     & 11453  &  1 & 7 & 0.043 & 5.7  \\
93.48     & 10697  &  0 & 6 &  --  & --   \\
94.09     & 10626  &  2 & 6 & 0.042 & 11.6 \\
97.43     & 10258  &  1 & 6 & 0.043 & 2.4  \\
102.9     &  9716  &  0 & 5 &  --  & --   \\
104.8     &  9545  &  2 & 5 & 0.040 & 23.6 \\
111.7     &  9418  &  1 & 5 & 0.043 & 2.0  \\
115.7     &  8642  &  0 & 4 &  --  & --   \\
116.7     &  8567  &  2 & 4 & 0.041 & 17.8 \\
125.9     &  7943  &  1 & 4 & 0.043 & 3.0 \\
134.5     &  7438  &  0 & 3 &  --  & --   \\
135.8     &  7364  &  2 & 3 & 0.039 & 38.6 \\
142.0     &  7041  &  1 & 3 & 0.043 & 0.9  \\
147.0     &  6804  &  0 & 2 &  --  & --   \\
147.4     &  6784  &  2 & 2 & 0.038 & 103  \\
173.5     &  5765  &  2 & 1 & 0.043 & 17.7 \\
174.9     &  5716  &  1 & 2 & 0.043 & 0.2  \\
176.3     &  5673  &  0 & 1 &  --  & --   \\
192.8     &  5188  &  2 & 0 & 0.034 & 210  \\
\enddata

\tablenotetext{e}{
$M=0.48020 M_{\odot}$, $M_H=0.00042 M_{\odot}$\\
$\log g = 5.72$, $T_{\rm eff}$ = 33,334 K \\
$v_{\rm rot}$ = 0.030 km/s, 
$1/P_{\rm surf}$ = 0.043 $\mu$Hz
}
\end{deluxetable}

%Table 7 - Old model, case D

\begin{deluxetable}{rrcccc}
\tabletypesize{\small}
\tablecaption{Representative post-EHB model\tablenotemark{f} 
pulsations, Case D}
\tablewidth{0pt}

\tablehead{
\colhead{Period} & \colhead{$\nu$ [$\mu$Hz]} & \colhead{$\ell$}  & 
\colhead{$n$} & \colhead{$1/P_{\rm surf} (1-C_{n,\ell})$}
&\colhead{$\Delta \nu$ [$\mu$Hz]}
}

\startdata
85.23     & 11800  &  0 & 7 &  --  & --   \\
85.46     & 11701  &  2 & 7 & 13.8 & 28.4 \\
88.48     & 11302  &  1 & 7 & 13.9 & 39.6 \\   
94.79     & 10257  &  0 & 6 &  --  & --  \\  
95.73     & 10444  &  2 & 6 & 13.5 & 33.0 \\
99.48     & 10052  &  1 & 6 & 14.1 & 30.7 \\
104.0     &  9615  &  0 & 5 &  --  & -- \\
105.7     &  9463  &  2 & 5 & 12.9 & 38.6 \\
113.8     &  8791  &  1 & 5 & 14.0 & 36.5 \\
118.1     &  8467  &  0 & 4 &  --  & -- \\  
118.9     &  8414  &  2 & 4 & 13.3 & 28.6 \\
126.7     &  7895  &  1 & 4 & 13.9 & 46.1 \\
135.8     &  7364  &  0 & 3 &  --  & -- \\
137.8     &  7257  &  2 & 3 & 12.4 & 49.7 \\
144.9     &  6902  &  1 & 3 & 14.0 & 26.3 \\ 
148.0     &  6758  &  2 & 2 & 12.9 & 64.5 \\
148.2     &  6748  &  0 & 2 &  -- & -- \\
177.5     &  5634  &  2 & 1 & 14.1 & 30.3 \\
179.3     &  5577  &  1 & 2 & 14.0 & 21.1 \\ 
180.7     &  5534  &  0 & 1 &  -- & -- \\
194.0     &  5154  &  2 & 0 & 11.3 & 137.7 \\
\enddata

\tablenotetext{f}{
$M=0.48010 M_{\odot}$, $M_H=0.00047 M_{\odot}$\\
$\log g = 5.70$, $T_{\rm eff}$ = 33,116 K \\
$v_{\rm rot}$ = 10 km/s, 
$1/P_{\rm surf}$ = 14.1 $\mu$Hz, $J/J_{\rm init}=0.076$ 
}
\end{deluxetable}

\begin{thebibliography}

%Behr et al. M15
\bibitem[Behr et al.(2000a)]{BehrM15}
Behr, B.B., Cohen, J.G., \& McCarthy, J.K. 2000a, \apjl, 531, L37

% Behr et al. M13
\bibitem[Behr et al.(2000b)]{BehrM13}
Behr, B.B., Djorgovski, S.G., Cohen, J.G., McCarthy, J.K., Cote, P.,
Piotto, G., \& Zoccali, M. 2000b, \apj, 528, 849

% Behr big paper on GCs
\bibitem[Behr(2003a)]{BehrGC}
Behr, B.B. 2003a, \apjs, 149, 67

% Behr big paper on field HB stars
\bibitem[Behr(2003b)]{BehrField}
Behr, B.B. 2003b, \apjs, 149, 101

% Brassard et al. 2001 PG 0014 tronta
\bibitem[Brassard et al.(2001)]{Brass0014}
Brassard, P., Fontaine, G., Bill\`eres, M., Charpinet, S., Liebert, J., 
\& Saffer, R. 2001, \apj, 563, 1013

%Brown et al. blue hook obs/theory
\bibitem[Brown et al.(2001)]{brownetal01}
Brown, T.M., Sweigart, A.V., Lanz, T., Landsman, W.B., \& Hubeny, I. 2001,
\apj, 562, 368

%Catellanin et al - semiconvection treatment
\bibitem[Castellani et al.(1985)]{Castetal85}
Castellani, V., Chieffi, A., Pulone, L., \& Tornamb\'e, A. 1985, \apj, 296, 204

%Caughlan and Fowler - final nuclear rates
\bibitem[Caughlan \& Fowler(1988)]{Caufow88}
Caughlan, G.R. \& Fowler, W.A. 1988, Atomic Data Nuc. Data Tables, 40, 284

%Charpinet et al. PASP sdBV review
\bibitem[Charpinet et al.(2001)]{Chaetal01}
Charpinet, S., Fontaine, G., \& Brassard, P. 2001, \pasp, 113, 775

%Charpinet et al. Paper II
\bibitem[Charpinet et al.(2002)]{CharpII}
Charpinet, S., Fontaine, G., Brassard, P., \& Dorman, B. 2002, \apjs, 139,
487

%Cox stellar pulsation book
\bibitem[Cox(1980)]{Cox80}
Cox, J.P. 1980, Stellar Pulsation (Princeton: Princeton University Press)

%D'Cruz et al - mass loss and EHB stars
\bibitem[D'Cruz et al.(1996)]{dcruz}
D'Cruz, N., Dorman, B., Rood, R., \& O'Connell, R. 1996, \apj, 466, 359

%Dehner and Kawaler - DB Diffusion
\bibitem[Dehner \& Kawaler(1995)]{Dehkaw95}
Dehner, B.T. \& Kawaler, S.D. 1995, \apj, 445, L141

%Dehner Ph.D.
\bibitem[Dehner(1996)]{Dehn96}
Dehner, B.T. 1996, Ph.D. dissertation, Iowa State University

%Deupree helium core flash
\bibitem[Deupree(1996)]{Deup96}
Deupree, R. 1996, \apj, 471, 377

%Green et al - Betsy star discovery paper
\bibitem[Green et al.(2003)]{Betsy03}
Green, E.M. et al. 2003, \apjl, 583, L31

%Han et al - binaries
\bibitem[Han et al.(2003)]{hanetal03}
Han, Z., Podsiadlowski, P., Maxsted, P., \& Marsh, T. 2003, \mnras, 341, 669

%Hansen and Kawaler
\bibitem[Hansen \& Kawaler(1994)]{hankaw}
Hansen, C.J. \& Kawaler, S.D. 1994, Stellar Interiors: Principles, Structure,
and Evolution, (New York: Springer)

%Heber et al - PG1605 rotation
\bibitem[Heber et al.(1999)]{hebetal99}
Heber, U., Reid, I.N., \& Werner, K. 1999, \aa, 348, L25

%Iglesias and Rogers latest OPAL (1996)
\bibitem[Iglesias \& Rogers(1996)]{Igrog96}
Iglesias, C.A. \& Rogers, F.J. 1996, \apj, 464, 943

%Kawaler Kraft Kurve
\bibitem[Kawaler(1987)]{Kaw87}
Kawaler, S.D. 1987, \pasp, 99, 1322

%Kawaler DB radial
\bibitem[Kawaler(1993)]{Kaw93}
Kawaler, S.D. 1993, \apj, 404, 294

%Kawaler and Bradley - nuff said
\bibitem[Kawaler \& Bradley(1994)]{Kawbra94}
Kawaler, S.D. \& Bradley, P.A. 1994, \apj, 427, 415

%Kawaler PG1605
\bibitem[Kawaler(1998)]{Kaw98}
Kawaler, S.D. 1998, in 11th European Workshop on White Dwarfs, ed. J.-E.
Solheim \& E. Meistas, (San Francisco: ASP), 158.

%Kawaler et al rotational inverson
\bibitem[Kawaler et al.(1999)]{KSG99}
Kawaler, S.D., Sekii, T., \& Gough, D. 1999, \apj, 516, 349

%Kawaler et al - Naples rotation paper
\bibitem[Kawaler et al.(2003)]{Kawetal03}
Kawaler, S.D., Hostler, S.R., \& Burkett, J. 2003, in NATO Advanced 
Research Workshop on White Dwarfs: the Thirteenth European Workshop, 
ed. R. Silvotti \& D. de Martino (Kluwer: Dordrecht)
 
%Kilkenny PG 1605
\bibitem[Kilkenny et al.(1999)]{Kilk1605}
Kilkenny, D., Koen, C., O'Donoghue, D., Van Wyk, F., Larson, K.A., 
Shobbrook, B., Sullivan, D.J., Burleigh, M., Dobbie, P., \& 
Kawaler, S.D. 1999, \mnras, 303, 525

% Kilkenny PG 1047
\bibitem[Kilkenny et al.(2001)]{Kilk1047}
Kilkenny, D., Bill\`eres, M., Stobie, R.S., Fontaine, G., Shobbrook, R.R.,
O'Donoghue, D., Brassard, P., Sullivan, D., Burleigh, M.R., \& Barstow, M.A.
2002, \mnras, 331, 399

% Kilkenny sdBV review
\bibitem[Kilkenny(2002)]{Kilk02}
Kilkenny, D. 2002, in Radial and Nonradial Pulsations as Probes of Stellar
Physics, ed. C. Aerts \& T. Bedding (San Francisco, ASP Press), p. 356

% Mengel et al - sdB and rotation
\bibitem[Mengel et al.(1976)]{mengetal76}
Mengel, J.G., Norris, J., \& Gross, P. 1976, \apj, 204, 488

%O'Brien & Kawaler - PG0122 Neutrinos
\bibitem[O'Brien \& Kawaler(2000)]{OBrkaw00}
O'Brien, M.S. \& Kawaler, S.D. 2000, \apj, 539, 372

%%Pesnell rotational splitting
%\bibitem[Pesnell(1985)]{Pes85}
%Pesnell, W.D. 1985, \apj, 292, 238

%Peterson initial paper
\bibitem[Peterson(1983)]{pete83}
Peterson, R.C. 1983, \apj, 275, 537

%Peterson next papers
\bibitem[Peterson(1985a)]{pete85a}
Peterson, R.C. 1985a, \apj, 289, 320

\bibitem[Peterson(1985b)]{pete85b}
Peterson, R.C. 1985b, \apj, 294, L35

%Petersen  rotation later work
\bibitem[Peterson et al.(1995)]{Petetal95}
Peterson, R.C., Rood, R.T., \& Crocker, D.A.  1995, \apj, 453, 214

%Peterson rotation in RR Lyr stars
\bibitem[Peterson et al.(1996)]{petetal96}
Peterson, R.C., Carney, B.W., \& Lathank D.W. 1996, \apjl, 465, L47

%P, D, D, - Yale models of RGB - HB rotation
\bibitem[Pinsonneault et al.(1991)]{Pinetal91}
Pinsonneault, M. H., Deliyannis, C. P., \& Demarque, P. 1991, \apj, 367, 239

%Recio-Blanco and EHB rotation
\bibitem[Recio-Blanco et al.(2002)]{Recetal02}
Recio-Blanco, A., Piotto, G., Aparicio, A., \& Renzini, A. 2002, \apjl,
572, L71

%Recio-Blanco and EHB rotation, II
\bibitem[Recio-Blanco et al.(2004)]{Recetal04}
Recio-Blanco, A., Piotto, G., Aparicio, A., \& Renzini, A. 2004, \aap, 
417, 597

%Mike Reed - Feige 48
\bibitem[Reed et al.(2004)]{ReedetalF48}
Reed, M. et al. (the WET collaboration) 2004, \mnras, 348, 1164

%Sandquist and Taam - CE binaries
\bibitem[Sandquist et al.(2000)]{sandtaam}
Sandquist, E., Taam, R., \& Burkert, A. 2000, \apj, 533, 984

% Sills \& Pinsonneault
\bibitem[Sills \& Pinsonneault(2000)]{SP2000}
Sills, A. \& Pinsonneault, M. 2000, \apj, 540, 489

%Soker and Harpaz planet spinup
\bibitem[Soker \& Harpaz(2000)]{SokHar00}
Soker, N. \& Harpaz, A. 2000, \mnras, 317, 861

%Stobie et al. - EC10228
\bibitem[Stobie et al.(1997)]{Stoetal97}
Stobie, R.S., Kawaler, S.D., Kilkenny, D., O'Donoghue, D., \& Koen, C. 1997,
\mnras, 285, 651

%Sweigart and Gross HB model construction
\bibitem[Sweigart \& Gross(1976)]{Swegro76}
Sweigart, A. \& Gross, P. 1976, \apjs, 32, 367

%Sweigart late flashers
\bibitem[Sweigart (1997)]{swei97}
Sweigart, A. 1997, in The Third Conference on Faint Blue Stars, ed. A.G.D.
Philip, J. Liebert, \& R.Saffer (Schenectady: L. Davis Press), 3

%Thompson et al GONG Science - rotation
\bibitem[Thompson et al.(1996)]{Thometal96}
Thompson, M.J. et al. 1996, Science, 272, 1300

%Unno et al. NRP book
\bibitem[Unno et al.(1989)]{Unetal89}
Unno, W., Osaki, Y., Ando, Y., Saio, H., \& Shibahashi, H. 1989,
Nonradial Oscillations of Stars, (Tokyo: Tokyo University Press)

%van Hoolst et al - nonlinear RR Lyra
\bibitem[van Hoolst et al.(1999)]{vanhetal98}
van Hoolst, T., Dziembowski, W.A., \& Kawaler, S.D. 1998, \mnras, 297, 536

%Wood et al - AGB/Mira nrp?
\bibitem[Wood et al.(2004)]{Woodetal04}
Wood, P.R., Olivier, E.A., \& Kawaler, S.D. 2004, \apj, 604, 800

%Zahn 92 - Jean-Paul-Zahn...
\bibitem[Zahn(1992)]{Zahn92}
Zahn, J.--P. 1992, \aap, 265, 115

\end{thebibliography}
\end{document}